\documentclass[twocolumn,showpacs,pra]{revtex4}
\usepackage{amssymb}
\usepackage{graphicx}
\usepackage{dcolumn}
\usepackage{bm}
\usepackage{amsmath}
\usepackage{epsfig}

\begin{document}

\title{Enhanced Molecular Orientation Induced by Molecular Anti-Alignment}

\author{E. Gershnabel}
\author{I. Sh. Averbukh}
\affiliation{Department of Chemical Physics, The Weizmann Institute
of Science, Rehovot 76100, ISRAEL}
\author{Robert J. Gordon}
\affiliation{Department of Chemistry, University of Illinois at
Chicago, Chicago, IL 60680-7061, USA}

\begin{abstract}

We explore the  role of laser induced anti-alignment in enhancing
molecular orientation. A field-free enhanced orientation via
anti-alignment scheme is presented, which combines a linearly
polarized femtosecond laser pulse with a half-cycle pulse. The laser
pulse induces transient anti-alignment in the plane orthogonal to
the field polarization, while the half-cycle pulse leads to the
orientation. We identify two qualitatively different enhancement
mechanisms depending on the pulse order, and optimize their effects
using classical and quantum models both at zero and non-zero
temperature.

\end{abstract}

\maketitle

\section{Introduction}

The concept of molecular alignment refers to the angular
localization of the symmetry axis of a molecule. The degree of
alignment is measured by $\langle\cos^2\theta\rangle$, where
$\theta$ is the angle between the symmetry axis  and a specified
space-fixed axis, and $\langle\rangle$ denotes an average over the
molecular ensemble. A molecule is said to be aligned if its symmetry
axis lies along the space-fixed axis, whereas it is defined to be
anti-aligned  if its symmetry axis is perpendicular to the
space-fixed axis. In recent years, various applications of molecular
alignment have been proposed, including high harmonic generation
\cite{Velotta}, laser pulse compression \cite{Bartels},
nanolithography \cite{Gordon}, control of photodissociation and
photoionization \cite{Larsen}, and quantum information processing
\cite{Shapiro}. These applications have stimulated the need to
 align  molecules optimally under field-free conditions. An important
milestone in the development of alignment methods is the use of
linearly polarized, ultrashort laser pulses to create a rotational
wave packet by an impulsive Raman mechanism. If the temporal  pulse
width of the laser is shorter than the rotational period of the
molecule (i.e., if its bandwidth is greater than the rotational
level spacing), the molecule undergoes a series of Raman excitations
that produce a coherent superposition of rotational states
\cite{Heritage}. For short pulses, peak field-free alignment along
the electric vector of the laser field is achieved after termination
of the laser pulse, at a time that depends on the pulse strength. As
the wave packet evolves, the molecule loses its alignment, and even
becomes $anti-aligned$ at some later time. The molecule also
undergoes a series of field-free realignments \cite{Rosca} at
integer multiples of the revival time, $\tau_{rev}=1/(2Bc)$, where
$B=\hbar/(4\pi c I_m)$ is the rotational constant, $c$ is speed of
light, and  $I_m$ is the moment of inertia. In addition, a number of
fractional rotational revivals occur at rational fractions of
$\tau_{rev}$ \cite{AP,Robinett}.

Molecular orientation refers to the case of molecules with a
directional symmetry axis (i.e., when a molecule has a permanent
dipole). Orientation is measured by $\langle\cos\theta\rangle$,
where $\theta$ is the angle between the molecular dipole and the
same spaced-fixed axis used to define alignment. Naturally,
orienting a molecule requires breaking of the orienting field
symmetry. Various symmetry-breaking methods have been proposed,
including introduction of a weak DC electric (or magnetic) field in
conjunction with a pulsed laser field \cite{Friedrich}, and coherent
excitation with laser fields of frequencies $\omega$ and $2\omega$
\cite{Charron}. The most versatile method for  orientating dipolar
molecules utilizes asymmetric electromagnetic half-cycle pulses
(HCPs) \cite{Dion,Averbukh}. Maximum field-free orientation is
achieved some time after  application of the HCP. A single short HCP
has a limited effect, however. As was shown in ref. \cite{Averbukh},
its effect saturates with intensity. An HCP applied to a group of
molecules that are initially oriented randomly in space  contributes
different angular velocities to individual molecules, so that
molecules starting from obtuse angles $(\pi/2\leq\theta\leq\pi)$
move too slowly to catch up with molecules starting from acute
angles $(0\leq\theta\leq\pi/2)$. As a result, the kicked molecules
do not all point in the same final direction at any given time, and
the orientation parameter has a maximum value of
$<\cos\theta>\approx0.75$. This effect is similar to non-perfect
focusing caused by spherical aberration in geometrical optics. It
was shown theoretically \cite{Averbukh,Leibscher,Sugny} that the
degree of field-free orientation and alignment can be enhanced by
using trains of laser pulses, and enhanced alignment by a pair of
pulses has been demonstrated experimentally
\cite{Lee,Bisgaard,Bisgaard1,Pinkham}. (For a recent review of
field-free alignment, see Refs. \cite{Stapelfeldt,review2}.)
Creating a train of HCPs with sufficient strength to orient a
molecule is experimentally difficult, however, requiring that
alternative approaches be developed.

As we recently showed \cite{Gershnabel}, anti-aligning molecules in
a plane perpendicular to the HCP polarization can enhance the
orienting effect of the HCP.  In the current paper we investigate in
detail  the "orientation via anti-alignment" approach that combines
an asymmetric half-cycle pulse with a symmetric femtosecond laser
pulse. The latter induces $anti-alignment$, whereas the HCP orients
the molecules. Depending on the pulse order, we identify two
different mechanisms for enhanced field-free orientation, and
describe them qualitatively  in section \ref{Mechanism}. A  rigorous
formulation of the problem is given in \ref{Problem}. A classical
approach to the model at zero and finite temperatures is presented
in section \ref{Classical}, and a full quantum treatment is
developed in section \ref{Quantum}, where we also demonstrate
further enhancement via a $three$ pulse scheme. Finally, we
summarize our findings  in section \ref{Conclusions}.

\section{"Orienting an anti-aligned state" and "Correcting the rotational velocity aberration"} \label{Mechanism}
 We identified two qualitatively different mechanisms for the
orientation enhancement, which we term "orienting an anti-aligned
state" and "correcting the rotational velocity aberration,"
depending on the temporal order of the pulses.

\begin{figure}[htb]
\begin{center}
\includegraphics[width=4cm]{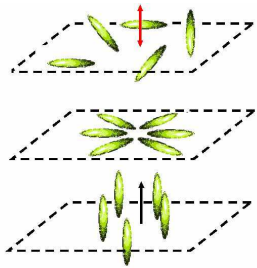}
\end{center}
\caption{"Orienting an anti-aligned state" mechanism. (Top) The
symmetric pulse is applied to an ensemble of randomly oriented
molecules. (Middle) After a time $t_1$, when the molecules are
anti-aligned,  an asymmetric orienting pulse is applied. (Bottom)
After an additional time $t_2$, the molecules are oriented.}
\label{Orienting an anti-aligned state}
\end{figure}

The "orienting an anti-aligned state" mechanism is illustrated in
Fig. \ref{Orienting an anti-aligned state}. The top drawing depicts
a randomly oriented collection of molecules. A short symmetric laser
pulse applied to the molecules pushes their  symmetry axes toward
the plane perpendicular to the desired orientation direction,
preparing the molecules in an $anti-aligned\:state$ angularly
localized near $\theta=\pi/2$. A schematic illustration of the
$anti-aligned\:state$ is given in the middle drawing. When a delayed
strong asymmetric HCP is applied to such an ensemble, all the
molecules gain nearly the same rotational velocity. Because all the
molecules depart from the anti-aligned state with very similar
initial angles, they reach the orientation direction almost
simultaneously at some later time, pointing in the same direction,
as depicted in the bottom drawing.
As is apparent  from the bottom two drawings in Fig. \ref{Orienting
an anti-aligned state}, the more strongly the molecules are
anti-aligned, the better they subsequently become oriented.

\begin{figure}[htb]
\begin{center}
\includegraphics[width=2cm]{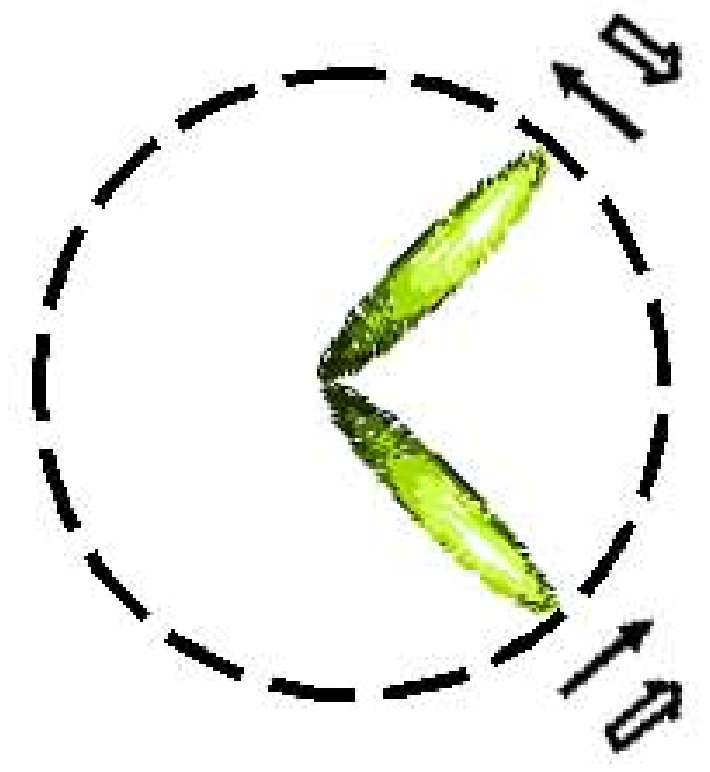}
\end{center}
\caption{"Correcting the rotational velocity aberration" mechanism.
Two schematic molecules are plotted. The filled arrows denote the
first HCP orienting direction, and the open arrows denote the second
symmetric pulse directions. The molecule at $\theta<\pi/2$ is
decelerated by the symmetric pulse, whereas the molecule at
$\theta>\pi/2$ is accelerated by the symmetric pulse.}
\label{Correcting the rotational velocity aberration}
\end{figure}

The "correcting the rotational velocity aberration" mechanism is
illustrated in Fig. \ref{Correcting the rotational velocity
aberration}. First, we apply the asymmetric HCP. The molecules gain
angular velocity in the direction illustrated in Fig.
\ref{Correcting the rotational velocity aberration} by the filled
arrows. Then, shortly $after$ the orienting HCP (or even
simultaneously with it), we apply the symmetric laser pulse. Such a
pulse decelerates the rotation of molecular dipoles having acute
angles with respect to the orientation direction and accelerates
those dipoles having obtuse angles. The directions of the angular
velocity induced by the symmetric pulse is illustrated in Fig.
\ref{Correcting the rotational velocity aberration} by the open
arrows. This effect compensates for  "spherical aberration" in the
angular distribution of the rotational velocity distribution and
improves the overall orientation at a later time.

In the following sections we  demonstrate for both mechanisms that
significantly enhanced orientation may be achieved by a proper
choice of the delay between the pulses and of their relative
intensities.

\section{Formulation of the Problem}\label{Problem}

The Hamiltonian of a 3D driven rigid rotor (linear molecule)
interacting with a linearly polarized field is given by
\begin{equation}
H=\frac{\hat{J}^{2}}{2I_m}+V(\theta,t), \label{hamiltonian}
\end{equation}
where $\hat{J}$ is the angular momentum operator and $\theta$ is the
angle between the molecular axis and the polarization vector of the
field. For a symmetric laser pulse interacting with the induced
polarization, the interaction term, averaged over the fast optical
oscillations, is given by
\begin{equation}
V(\theta,t)=-\frac{1}{4}\varepsilon^{2}(t)[(\alpha_{\|}-\alpha_{\bot})\cos^{2}(\theta)+\alpha_{\bot}],
\label{inducedint}
\end{equation}
where $\varepsilon(t)$ is the envelope of the laser field, and
$\alpha_{\|}$ and $\alpha_{\bot}$ are the parallel and perpendicular
components of the polarizability tensor, respectively. For a
symmetric laser pulse the contribution from the permanent dipole
averages to zero.

For an asymmetric HCP, the interaction with the dipole moment is
given by
\begin{equation}
V(\theta,t)=-\mu\varepsilon(t)\cos(\theta), \label{dipole}
\end{equation}
where $\mu$ is the permanent dipole moment and $\varepsilon(t)$ is
the amplitude of the HCP. In the present paper we assume that the
duration of the laser pulse is much shorter than the contributing
periods of the rotational wave packet, so that the excitation
dynamics may be calculated in the impulsive limit. The impulse
imparted to the rotator is characterized by a dimensionless kick
strength (or action), P. For an asymmetric pulse, P is given by
\begin{equation}
P_a=(\mu/\hbar)\int_{-\infty}^{\infty}\varepsilon(t)dt, \label{Pa}
\end{equation}
where the integration is performed over the  unidirectional part of
the HCP, and
\begin{equation}
P_s=(1/4\hbar)(\alpha_{\|}-\alpha_{\bot})\int_{-\infty}^{\infty}\varepsilon^2(t)dt
\label{Ps}
\end{equation}
for a symmetric pulse. We start with the mechanism of "orienting an
anti-aligned state" and consider a rotator excited first with a
symmetric pulse of strength $P_s$ at $t=0$ and then with an
asymmetric pulse of strength $P_a$ at $t=t_1$. Henceforth the
dimensionless time is measured in the units of
$I_m/\hbar=\tau_{rev}/2\pi$.

\section{Classical Treatment}\label{Classical}

Considerable physical insight may be derived from the
(semi)-classical treatment of the problem, which is valid for $P_s,
P_a \gg 1$. This is a natural approach to the process of enhanced
orientation involving highly excited rotational states. In this
section we formulate the orientation problem classically. A more
general (although less intuitive) quantum mechanical treatment of
the same problem is provided in section \ref{Quantum}. Classically,
if a rotationless molecule is initially aligned at angle $\theta_0$,
it will be found at the same angle just after the first symmetric
kick (see Eqs.(\ref{inducedint}),(\ref{Ps})) but with angular
velocity $-P_s\sin(2\theta_0)$, so that at some later time it will
have an angle
\begin{equation}
\theta(t)=\theta_0-P_s t\sin(2\theta_0). \label{First Angle}
\end{equation}
When the orienting pulse of strength $P_a$ is applied at time
$t=t_1$, the velocity increment is $-P_a\sin[\theta (t_1)]$, so that
the angular velocity after the second pulse is
$\omega(\theta_0)=-P_s\sin(2\theta_0)-P_a\sin[\theta_0-P_st_1\sin(2\theta_0)]$.
The angle $\theta$ at time $t_2$ after the second pulse is therefore
\begin{eqnarray}
\theta(t_1+t_2)&=&\theta_0-P_st_1\sin(2\theta_0)-t_2\{P_s\sin(2\theta_0)\nonumber\\
&+&P_a\sin[\theta_0-P_st_1\sin(2\theta_0)]\}. \label{Second Angle}
\end{eqnarray}

A similar expression may be derived for the inverse order of pulses,
utilized in the second mechanism of enhanced orientation. The
alignment at time $t=t_1$ and orientation at time $t=t_1+t_2$ is
calculated by averaging $\cos^k\theta(t)$ over all values of
$\theta_0$,
\begin{equation}
\langle\cos^k\theta(t)\rangle=\frac{1}{2}\int_0^{\pi}\cos^k\theta(t)\sin\theta_0d\theta_0,
\label{Average}
\end{equation}
where $k=1$ and $2$ for orientation and alignment, respectively.

\begin{figure}[htb]
\begin{center}
\includegraphics[width=8cm]{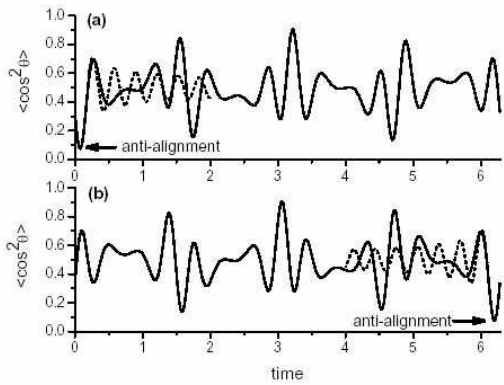}
\end{center}
\caption{Alignment parameter vs time after excitation by a laser
pulse with (a) $P_s=-10$ and (b) $P_s=10$. Solid curves are the
quantum results. Dashed curves are calculated classically (a) for
positive time and (b) for negative time (shifted by $\tau_{rev}$).
Strong anti-alignment is seen in both cases.} \label{AlignmentGraph}
\end{figure}

Our analysis shows that strong transient anti-alignment may be
achieved via two related methods. The first one is of a classical
nature. It operates on a short time-scale ($t<<\tau_{rev}$) and
requires $negative$ values of the kick strength $P_s$. Pulses with
negative $P_s$ produce anti-alignment by pushing molecules into the
equatorial plane ($\theta=\pi/2$). Fig. \ref{AlignmentGraph}a shows
the expectation value of $\langle\cos^2\theta\rangle$ calculated
both classically and quantum mechanically (as will be explained in
section \ref{Quantum}) for a pulse with $P_s=-10$. Both treatments
predict a deep minimum ($\langle\cos^2\theta\rangle_{min}=0.077$)
shortly after the pulse at $t_m\approx 0.8/|P_s|$. There are various
ways of achieving a $negative$ kick strength of the symmetric pulse.
First, some alkali halides (such as $LiF$) have a negative
polarizability anisotropy ($\alpha_{\|}-\alpha_{\bot}$)\cite{Pluta}.
Second, the interaction of molecules with a circularly polarized
light pulse propagating in the direction of the desired orientation
is proportional to $P_s\sin^2\theta=P_s-P_s\cos^2\theta$. This
result is formally equivalent to the interaction with a linearly
polarized pulse (our model) having negative kick strength.

A third method of achieving anti-alignment uses laser pulses with
$positive$ $P_s$, which cause a substantial $alignment$ on a short
(classical) time scale $after$ the kick (see Fig.
\ref{AlignmentGraph}b). However, if it were possible to invert the
dynamics and travel backwards in time, one would observe a strong
anti-alignment $before$ the kick. Remarkably, the effect of quantum
revivals \cite{Robinett} provides such an option. Indeed, the
rotational wave function satisfies the following equation
\begin{equation}
\Psi(\theta,\tau_{rev}-t)=\Psi(\theta,-t), \label{PsiNegTime}
\end{equation}
and, for sufficiently strong  pulses, quantum dynamics just before
one full revival cycle is equivalent to classical dynamics
analytically continued to $negative$  times. As a result,
considerable anti-alignment is observed in this time domain (see
Fig. \ref{AlignmentGraph}b).

\begin{figure}[htb]
\begin{center}
\includegraphics[width=8cm]{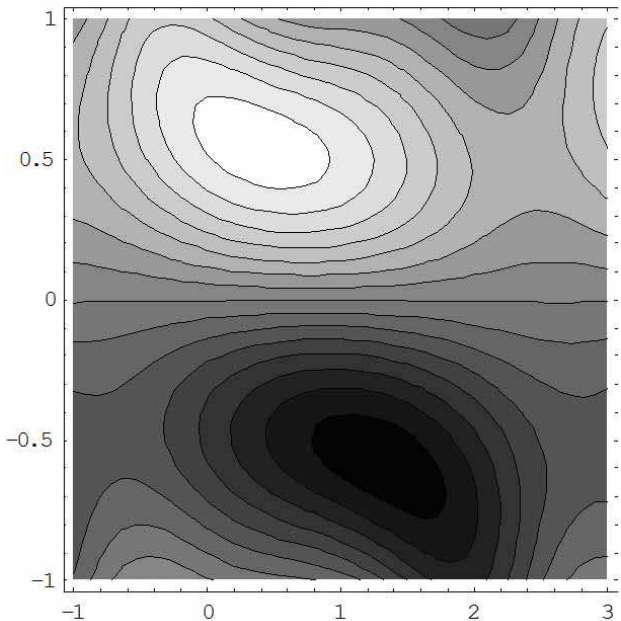}
\end{center}
\caption{Classical orientation factor as a function of $|P_s|t_1$
(horizontal axis) and $|P_s|t_2$ (vertical axis) for the first
mechanism (the laser pulse precedes the HCP). $P_a/|P_s|=3$. There
are two almost equivalent extreme points.}
\label{CountourAntiAlignment}
\end{figure}

Next we would like to optimize the orientation factor
$|\langle\cos(\theta)\rangle(P_s,P_a,t_1,t_2)|$ using the classical
model (Eqs. (\ref{Second Angle}) and (\ref{Average})) with an
anti-aligning pre-pulse ($P_s<0$). In this procedure, the model is
formally extended to negative times to cover effects in the full
revival time-domain, as  explained above. At zero initial
temperature, the optimal solution depends only on the ratio
$P_a/P_s$ and the products $P_st_1$ and $P_at_2$. A contour plot of
the orientation factor  as a function of $|P_s |t_1$ and $|P_s |t_2$
(for a fixed ratio $P_a/|P_s|=3$) is given in Fig.
\ref{CountourAntiAlignment}, where two almost equivalent extremum
points are observed.

\begin{figure}[htb]
\begin{center}
\includegraphics[width=8cm]{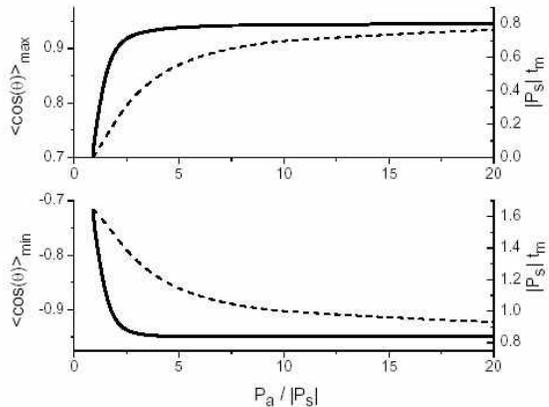}
\end{center}
\caption{Classically optimized orientation factor (solid curves) and
delay between pulses (dashed curves). The laser pulse is fired
before the half-cycle pulse.} \label{OrientationFactor}
\end{figure}

Figure \ref{OrientationFactor} displays the highest (optimized with
respect to the delay time) post-pulse orientation parameter,
$\langle\cos\theta\rangle$ (solid line), and the optimal delay
between pulses, $t_{1opt}$ (dashed line), as functions of
$P_a/|P_s|$. There are two optimal solutions of almost the same
efficiency. The first one provides maximal orientation in the
direction of the HCP shortly after the second pulse (upper panel in
Fig. \ref{OrientationFactor}). The second one (lower panel in Fig.
\ref{OrientationFactor}) delivers enhanced orientation in the
$opposite$ direction in the full revival domain after the second
pulse. In both cases, an impressive value of
$|\langle\cos\theta\rangle|\approx 0.95$ is achieved for rather
modest pre-pulses ($|P_s|\sim 0.1P_a$) (as compared to the limit of
$\langle\cos\theta\rangle_{max}\approx 0.75$ for a single HCP). In
this regime, the optimal delay between pulses, i.e. optimal $t_1$,
asymptotically approaches the time of the best anti-alignment,
$t_m\approx 0.8/|P_s|$.

It is easy to show that the same degree of orientation may be
achieved by combining an HCP with an aligning laser pulse ($P_s>0$).
The apparent symmetry relations,
\begin{equation}
\langle\cos\theta\rangle(P_s,P_a,-t_1,-t_2)=\langle\cos\theta\rangle(-P_s,-P_a,t_1,t_2)
\label{Symmetry Relation 1}
\end{equation}
and
\begin{equation}
\langle\cos\theta\rangle(P_s,P_a,t_1,t_2)=-\langle\cos\theta\rangle(P_s,-P_a,t_1,t_2)\label{Symmetry
Relation 2},
\end{equation}
reduce this problem to the already-studied case of the anti-aligning
pulse.

\begin{figure}[htb]
\begin{center}
\includegraphics[width=8cm]{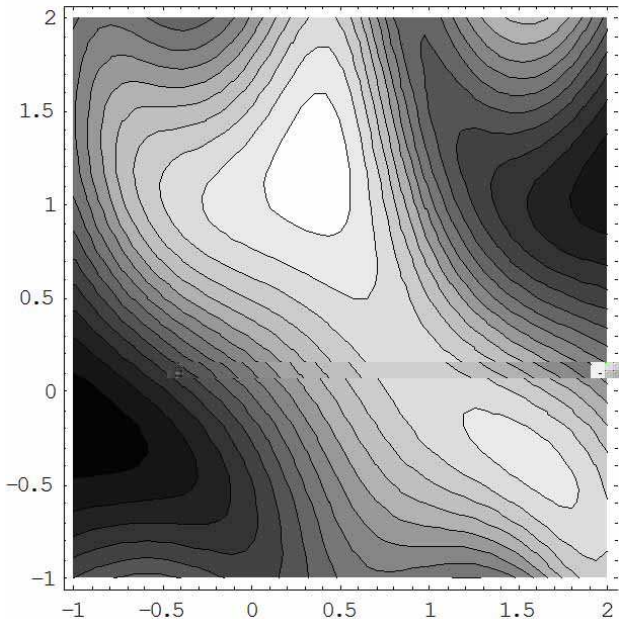}
\end{center}
\caption{Classical orientation factor as a function of $|P_s|t_1$
(horizontal axis) and $|P_s|t_2$ (vertical axis) for the second
orientation mechanism (HCP  precedes the laser pulse). $P_a/|P_s|=
1.6$. There is a single dominant maximum.}
\label{CountourAntiAlignment2}
\end{figure}

\begin{figure}[htb]
\begin{center}
\includegraphics[width=8cm]{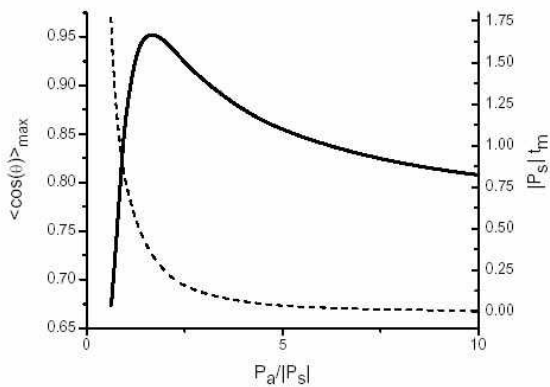}
\end{center}
\caption{Classically optimized orientation factor (solid curve) and
delay between pulses (dashed curve). The laser pulse is fired after
the half-cycle pulse.} \label{OrientationFactor2}
\end{figure}

We used the same approach to analyze the second mechanism of
enhanced orientation mentioned in the Introduction. In the simplest,
non-optimized version, the orienting and anti-aligning pulses are
applied simultaneously ($t_1=0$). Direct numerical maximization of
the expression in Eq. \ref{Average} shows that
$\langle\cos\theta\rangle_{max}=0.89$ when $P_a/|P_s|\approx 2.34$
and $|P_s|t_2\approx 0.78$. When the hybrid pulse is composed of an
orienting component and an $aligning$ one ($P_s>0$), the symmetry
relations (\ref{Symmetry Relation 1}) and (\ref{Symmetry Relation
2}) predict the same orientation, but in the opposite direction
($\langle\cos\theta\rangle=-0.89$) just before one full revival
cycle ($P_st_2\approx -0.78$). This effect was reported in a recent
paper \cite{Daems} as a result of direct numerical simulation of the
quantum rotational dynamics of molecules excited by a single hybrid
pulse. More efficient results may be obtained when the HCP precedes
the anti-aligning laser pulse. In this case, our classical analysis
reveals a single dominating optimal solution. A contour plot of the
orientation parameter is given in Fig. \ref{CountourAntiAlignment2}.
The optimal dominant solution is presented in Fig.
\ref{OrientationFactor2}. The maximum orientation
$\langle\cos\theta\rangle_{max}\approx 0.96$ is reached at
$P_a/|P_s|\approx 1.6$, and the optimal delay is
$t_m\approx0.36/|P_s|$.

As the next step, we calculate the orientation factor at a nonzero
temperature. We shall derive the orientation factor for the
"orientation of an anti-aligned state" mechanism only; the
orientation factor for the second mechanism may be similarly
derived.

The Lagrangian for the 3D free rotor is given by
\begin{equation}
L=\frac{1}{2}I_m(\dot{\theta}^2+\dot{\phi}^2\sin^2\theta),
\label{Lagrangian}
\end{equation}
and the Euler-Lagrange equation for $\theta$ is
\begin{equation}
\frac{d}{dt}\frac{\partial L}{\partial \dot{\theta}}-\frac{\partial
L}{\partial \theta}=0. \label{EulerLagrange}
\end{equation}
The canonical momenta are
\begin{equation}
P_{\phi}=\frac{\partial L}{\partial
\dot{\phi}}=I\dot{\phi}\sin^2\theta, \label{CanonicalMomentaPhi}
\end{equation}
(which is a constant of motion as $\phi$ is a cyclic coordinate),
and
\begin{equation}
P_{\theta}(t)=\frac{\partial L}{\partial
\dot{\theta}}=I\dot{\theta}. \label{CanonicalMomentaTheta}
\end{equation}
From Eqs. (\ref{EulerLagrange}), (\ref{CanonicalMomentaPhi}), and
(\ref{CanonicalMomentaTheta}) the  following equation of motion
follows:
\begin{equation}
\ddot{\theta}=\frac{P^2_{\phi}}{I^2}\frac{\cos\theta}{\sin^3\theta}.
\label{motionEq}
\end{equation}

In what follows, it is convenient to measure the canonical momenta
in the units of $p_{th}=I_m\omega_{th}$ with
$\omega_{th}=\sqrt{k_BT/I_m}$, where $T$ is the  temperature and
$k_B$ is the Boltzmann constant. By setting
$P'_{\phi}=P_{\phi}/p_{th}, P'_{\theta}=P_{\theta}/p_{th}$, and
$t'=\omega_{th}t$, one writes Eq. (\ref{motionEq}) as
\begin{equation}
\ddot{\theta}=P'^2_{\phi}\frac{\cos\theta}{\sin^3\theta}.
\label{NormmotionEq}
\end{equation}
The solution to this equation is
\begin{eqnarray}
\cos\theta(t')&=&\frac{1}{2}(1-\frac{P'_{\theta}}{\omega})\cos(\theta(0)-\omega
t')\nonumber \\
&+&\frac{1}{2}(1+\frac{P'_{\theta}}{\omega})\cos(\theta(0)+\omega
t') \label{cosTheta},
\end{eqnarray}
where
\begin{equation}
\omega=(P'^2_{\theta}+\frac{P'^2_{\phi}}{\sin^2\theta(0)})^{1/2}.
\end{equation}
Here $P'_{\theta}$  is a constant  initial canonical momentum (see
Eq. (\ref{CanonicalMomentaTheta})). As a result of a kick induced by
a symmetric laser pulse applied at $t=0$,  $P'_{\theta}$ changes its
value to
\begin{equation}
P'_{\theta}=P'_{\theta}(0)-P_s'\sin(2\theta(0)). \label{DipoleKick}
\end{equation}
The initial thermal distribution function for the molecules has the
following form in the chosen dimensionless variables:
\begin{equation}
f(\theta,\phi,P'_{\theta},P'_{\phi})=\frac{1}{8\pi^2}\exp[-\frac{1}{2}(P'^2_{\theta}+\frac{P'^2_{\phi}}{\sin^2\theta})]
\label{DistributionClassics}.
\end{equation}
The thermally averaged alignment factor reads as
\begin{eqnarray}
\langle\cos^2\theta\rangle(t')&=&\int_0^{\pi}
d\theta(0)\int_0^{2\pi}d\phi(0)\int_{-\infty}^{\infty}dP'_{\theta}(0)\int_{-\infty}^{\infty}dP'_{\phi}(0)\nonumber\\
&\times&\cos^2\theta(t')f(\theta(0),\phi(0),P'_{\theta}(0),P'_{\phi}(0)),
\label{AlignmentFactor}
\end{eqnarray}
where $f$ is given by Eq. (\ref{DistributionClassics}). The factor
$\cos\theta(t')$ is provided by Eq. (\ref{cosTheta}), in which
$P'_{\phi}=P'_{\phi}(0)$, and  $P'_{\theta}$  is given by Eq.
(\ref{DipoleKick}). An HCP is applied with a delay $t_1'$ after the
first symmetric laser pulse. Using the equations
$P'_{\theta}=\dot{\theta}$ and $d( \cos\theta )/ dt
=-\sin\theta\dot{\theta}$ we obtain $P'_{\theta}$ after the second
pulse:
\begin{equation}
P'_{\theta}=-\frac{d( \cos\theta )/
dt}{\sin\theta}-P'_{a}\sin\theta. \label{InducedDipoleKick}
\end{equation}
Here $\sin\theta$ and $\cos\theta$ are obtained from Eq.
(\ref{cosTheta}) at $t'=t_1'$. The new value of $P'_{\theta}$  and
the angle $\theta(t_1')$ are substituted again into Eq.
(\ref{cosTheta}) to propagate the rotation angle further. Thus, the
time-dependent orientation factor is given by
\begin{eqnarray}
\langle\cos\theta\rangle(t_1'+t_2')&=&\int_0^{\pi}
d\theta(0)\int_{0}^{2\pi}d\phi(0)\nonumber\\
&\times&\int_{-\infty}^{\infty}dP'_{\theta}(0)\int_{-\infty}^{\infty}dP'_{\phi}(0)\nonumber\\
&\times&\cos\theta(t'=t_1'+t_2')\nonumber\\
&\times&f(\theta(0),\phi(0),P'_{\theta}(0),P'_{\phi}(0)).
\label{OrientationFactorEq}
\end{eqnarray}
From here on, we return to the previous dimensionless  units used in
this paper. In particular, the dimensionless kick strengths are
replaced by
\begin{equation}
P'_{a,s}=P_{a,s}\frac{\hbar}{\sqrt{k_BTI_m}}\label{PRelation}.
\end{equation}

\begin{figure}[htb]
\begin{center}
\includegraphics[width=8cm]{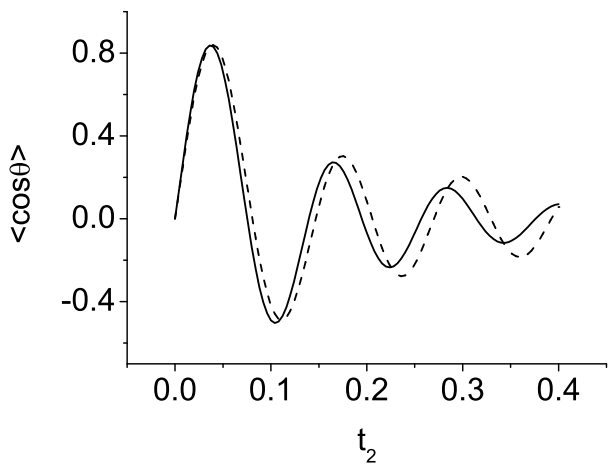}
\end{center}
\caption{Orientation parameter vs. $t_2$ ($t_1=0$) for $T=5K$,
$P_1=-10$, and $P_2=50$. Solid  and dashed lines are quantum and
classical calculations, respectively.} \label{ClassicalComparisonT}
\end{figure}

Figure \ref{ClassicalComparisonT} displays a comparison between
quantum (see the next section) and classical calculations for $KCl$
molecule kicked by two coinciding pulses ($t_1=0$)  at $T=5K$. One
may observe that the classical and quantum results agree well for
strong pulses ($P\gg1$) and short propagation time.

\section{Quantum Treatment} \label{Quantum}
Consider a quantum rotator being initially in the ground state and
kicked by a short symmetric laser pulse at $t=0$. In the impulsive
approximation, its wave function acquires a phase factor as a result
of the kick:
\begin{equation}
\Psi(\theta,t=+0)=\exp[iP_s\cos^2\theta]Y_0^0(\theta).\label{pss}
\end{equation}
By expanding this expression as a sum of spherical harmonics, one
finds the wave function at later time t:
\begin{equation}
\Psi(\theta,t)=\frac{1}{\sqrt{4\pi}}\sum_{J=0}c_J
\exp[-iJ(2J+1)t]Y_{2J}^0(\theta). \label{PsiQuantT0}
\end{equation}
The coefficients $c_J$ are given by \cite{Leibscher}
\begin{eqnarray}
c_J&=&\sqrt{\pi(4J+1)}(iP_s)^J\frac{\Gamma(J+1/2)}{\Gamma(2J+3/2)}\nonumber\\
&\times&_1F_1[J+1/2,2J+3/2,iP_s], \label{CoeffQuantT0}
\end{eqnarray}
where $_1F_1$ is the confluent hypergeometric function. At $t=t_1$
the rotator is kicked by the orienting pulse, acquiring an
additional phase factor
\begin{equation}
\Psi(\theta,t_1+0)=\exp[iP_a\cos\theta]\Psi(\theta,t_1-0).
\end{equation}
We use the well-known expression
\begin{equation}
\exp(iP_a\cos\theta)=\sum_{J=0}^{\infty}i^J\sqrt{4\pi(2J+1)}j_j(P_a)Y_j^0
(\theta),\label{Expression}
\end{equation}
where $j_j(P_a)$ is a spherical Bessel function, and again expand
the wave function in a series of spherical harmonics
\begin{equation}
\Psi(\theta,t_1+0)=\sum_{l=0}^{\infty}d_lY_l^0(\theta),
\end{equation}
where
\begin{eqnarray}
d_l&=&\frac{1}{\sqrt{4\pi}}\sum_{l'=0}^{\infty}\sum_{j=0}^{\infty}i^j\sqrt{(2J+1)}j_j(P_a)c_{l'}\exp[-il'(2l'+1)t_1]\nonumber\\
&\times&\sqrt{(2j+1)(4l'+1)(2l+1)}\frac{C(j,2l',l|0,0,0)^2}{2l+1}.\label{d_Expression}
\end{eqnarray}
Here $C(j,2l',l|0,0,0)$ is a Clebsch-Gordan coefficient. This new
wave function is allowed to propagate freely until $t=t_1+t_2$, at
which point the orientation and alignment parameters are calculated
by
\begin{equation}
\langle\cos^k\theta\rangle=\langle\Psi(\theta,t)|\cos^k\theta|\Psi(\theta,t)\rangle,k=1,2.
\end{equation}

\begin{figure}[htb]
\begin{center}
\includegraphics[width=8cm]{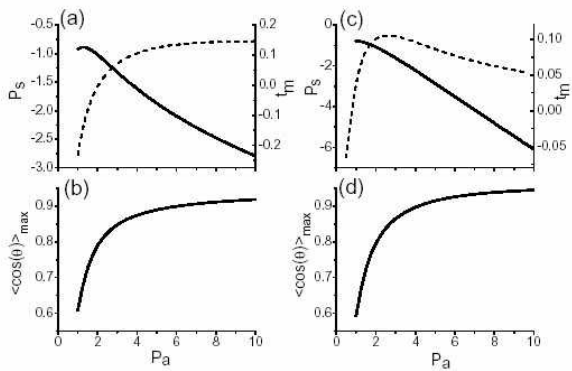}
\end{center}
\caption{Quantum mechanical optimized results at zero temperature.
The left column (a,b) corresponds to the mechanism of "orienting an
anti-aligned state" (laser pulse fired before the HCP). The right
column corresponds to the mechanism of "correcting the rotational
velocity aberration" (inverse order of pulses). Upper panels (a,c)
display the optimal strength of the anti-aligning pulse (solid
lines) and delay between pulses (dashed lines) as a function of the
HCP strength. Lower panels (b,d) present the highest value of the
post-pulse orientation factor vs the strength of the HCP.}
\label{OrientationQuantum}
\end{figure}

We performed a fully quantum-mechanical analysis of the "orientation
via anti-alignment" and "correcting the rotational velocity
aberration" mechanisms, using the above-described methodology.
Figure \ref{OrientationQuantum} demonstrates the optimized values of
the anti-aligning pulse strength, $|P_s|$, the optimum delay between
pulses, $t_m$, and the maximal orientation parameter,
$\langle\cos\theta\rangle_{max}$, as function of $P_a$. Very good
agreement between quantum and classical results is observed even for
moderate anti-aligning and orienting pulses ($P_s,P_a\sim 3$).
Remarkably, a significantly enhanced orientation may be achieved
with field strengths available currently in the laboratory.
Considering a $KCl$ molecule in the ground state (having a revival
time $t_{rev}\approx 128 ps$, a dipole moment $\mu\approx 10.3 D$,
and a polarization anisotropy $(\alpha_{\|}-\alpha_{\bot})\approx
3.1 {\AA}^3$, data taken from \cite{Pluta}), one expects $P_a\sim
10$ for an HCP with the amplitude of $85 kV/cm$ and a duration of
about $2ps$ ($1/e$ half width). According to Fig.
 \ref{OrientationQuantum}, the orientation factor
$\langle\cos\theta\rangle\approx 0.95$ may be observed if the HCP is
followed by a delayed anti-aligning pulse of $2ps$ duration and
$5\times10^{11}W/cm^2$ peak intensity. Negative times in Fig.
\ref{OrientationQuantum} correspond, as was explained above, to the
dynamics in the full revival domain.
\\
\\

\begin{figure}[htb]
\begin{center}
\includegraphics[width=8cm]{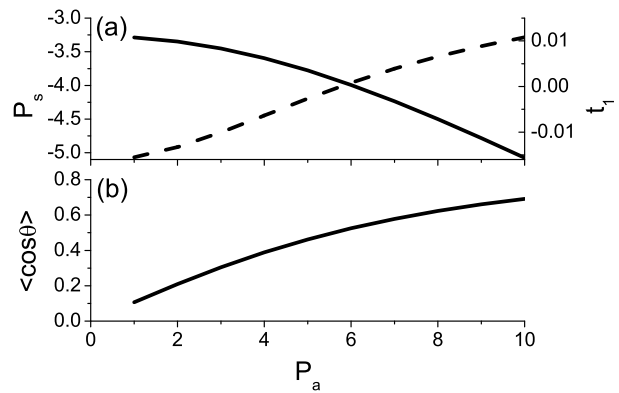}
\end{center}
\caption{Quantum mechanical optimized results for KCl at $T = 5K$.
The graphs correspond to the mechanism of orienting an anti-aligned
state (laser pulse fired before the HCP). a) Anti-aligning pulse
strength and delay between pulses vs. HCP strength are given by
solid and dashed lines, respectively. b) Orientation parameter vs.
HCP strength.} \label{2PulsesT}
\end{figure}

\begin{figure}[htb]
\begin{center}
\includegraphics[width=8cm]{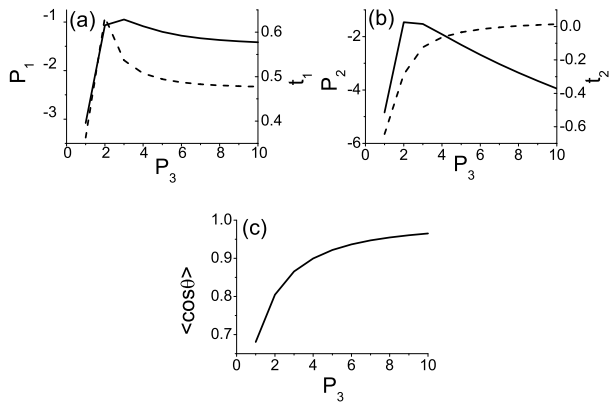}
\end{center}
\caption{Quantum mechanically optimized results for the three-pulse
mechanism, where the first two pulses are symmetric, and the third
pulse is an asymmetric HCP. a) First pulse strength $P_1$ and the
delay between first and second pulses $t_1$ vs. HCP strength are
given by solid and dashed lines, respectively. b) Second pulse
strength $P_2$ and the delay between second and third pulses $t_2$
vs. HCP strength are given by solid and dashed lines, respectively.
c) Orientation parameter vs. HCP strength.} \label{3pulses}
\end{figure}

We also analyzed quantum mechanically the orientation factor after
two-pulse excitation for an ensemble having a non-zero temperature.
The details of calculations can be found in the Appendix to this
paper. The optimized solution for the orientation of an anti-aligned
state is plotted in Fig. \ref{2PulsesT} as a function of $P_a$. A
comparison between the classical and quantum thermal results was
given in Fig. \ref{ClassicalComparisonT}.
One may observe in Fig. \ref{2PulsesT} that the temperature reduces
the orientation effect. However, the latter is still considerably
better compared to the single HCP excitation. The maximal
orientation factor for $KCl$ excited with a single HCP of strength
$P_a=10$  is $0.59$ at $T=5K$. The combined action of the HCP and
laser pulse provides a much higher value of $0.69$ at the same
thermal conditions. To achieve even better orientation, one may try
to add additional laser pulses to improve the anti-alignment in the
system. The simplest scheme of this kind consists of  two symmetric
laser pulses separated by a delay $t_1$, followed by an HCP after a
second delay, $t_2$. The optimal field-free orientation occurs at
some time $t_3$ after the HCP. A full quantum optimization of this
scheme at finite temperature presents a challenging numerical task,
and goes beyond the framework of the present paper. The results of
optimization at zero temperature are shown in Fig. \ref{3pulses}. In
particular, using the experimentally feasible  HCP discussed above
(i.e. $P_a=10$ at $85 kV/cm$) we obtain the optimized orientation
factor  $\langle\cos(\theta)\rangle=0.97$ for $P_{s1}=-1.42$ and
$P_{s2}=-3.95$. This means a 25\% reduction in the width of the
angular distribution compared to the  two-pulse orientation scheme.

\section{Conclusions} \label{Conclusions}

We presented an experimentally feasible method for enhanced
orientation of linear dipolar molecules using a half-cycle pulse
combined with a delayed laser pulse inducing molecular
anti-alignment. Two qualitatively different enhancement mechanisms
were identified depending on the pulse order, and their effects were
optimized with the help of quasi-classical as well as fully quantum
models. The transparent physics behind the interplay between
anti-alignment, orientation, and quantum rotational revivals
provides a solid basis for the future design of more sophisticated
and efficient solutions. In particular, we demonstrated that
enforced anti-alignment by a pair of symmetric laser pulses prior to
applying an HCP improves the orientation even more.  These results
may be generalized to  trains of multiple laser pulses, similar to
forced multi-pulse alignment techniques
\cite{Averbukh,Leibscher,LeibscherOther,Sugny,Lee,Bisgaard,Bisgaard1,Pinkham}.
In the same way, the "spherical aberrations" of a single HCP may be
better corrected by multiple series of delayed laser pulses as well.
Finally, a hybrid pair of delayed pulses may be followed by a series
of well-timed symmetric laser pulses designed to preserve  the
achieved orientation over an extended time period
\cite{Matos,Ortigoso}.

\section{Acknowledgments}
IA wishes to thank the Israel Science Foundation for support of this
research, and RJG acknowledges the Motorola Corporation and the US
Department of Energy for its support.

\section{Appendix}
We  present here  the finite-temperature results for the
"orientation of an anti-aligned state" mechanism only. The
orientation factor for the second orientation mechanism may be
derived similarly.

At $T>0,$ the orientation factor should be thermally averaged over
all initial states, so that
\begin{equation}
\langle\cos\theta\rangle=\sum_{l_0=0}^{\infty}
P(l_0)\sum_{m_0=-l_0}^{l_0}A^{m_0}_{l_0}(t), \label{Alignment}
\end{equation}
where
\begin{equation}
P(l_0)=\frac{1}{Q}\exp
[-\frac{l_0(l_0+1)}{2\sigma_{th}^2}]\label{Distribution}
\end{equation}
is the thermal distribution at temperature $T$,  $Q$ is the
rotational partition function, and $\sigma_{th}=(k_BT/2Bhc)^{1/2}$.
In Eq. (\ref{Alignment}),
\begin{equation}
A_{l_0}^{m_0}(t)=\langle\psi_{l_0}^{m_0}(t)|\cos(\theta)|\psi_{l_0}^{m_0}(t)\rangle\label{Alignmentl0m0}
\end{equation}
is the the contribution to the orientation parameter from initial
state $|l_0,m_0\rangle$.

The wave function  $|\psi_{l_0}^{m_0}(t)\rangle$, is calculated as
follows. After the first symmetric laser pulse, the wave function
becomes
\begin{equation}
|\psi_{l_0}^{m_0}(0^+)\rangle =
 \exp[iP_s\cos^2(\theta)]|l_0,m_0\rangle\label{induced}.
\end{equation}
Next, we expand $|\psi_{l_0}^{m_0}(0^+)\rangle$ in the spherical
harmonic basis,
\begin{equation}
\psi_{l_0}^{m_0}(0^+)=\sum_{l_0=0}^{\infty}\alpha_{l}^{m_0}|l,m_0\rangle\label{FirstExpansion},
\end{equation}
where the quantum number $m_0$ is preserved. The coefficient
$\alpha_l^{m_0}$ may be derived by expanding
$\exp[iP_s\cos^2(\theta)]$ in a sum of spherical harmonics using
Eqs.(\ref{pss}),(\ref{PsiQuantT0}), and (\ref{CoeffQuantT0}) and
applying the rules of angular momentum algebra. However, we employ
here another mathematically equivalent approach that is numerically
advantageous for large values of $P_s$. For this, we expand
$\exp[iP_s\cos^2(\theta)]$ in a series of Legendre polynomials,
using the formula similar to Eq. (\ref{Expression}), which gives
(for negative $P_s$):
\begin{eqnarray}
\exp[iP_s\cos^2(\theta)]&=&\exp[-i\frac{|P_s|}{2}]\sum_{J=0}^{\infty}i^{3J}(2J+1)j_J(\frac{|P_s|}{2})\nonumber\\
&\times&P_J(\cos(2\theta))\label{ExpansionNegPs}.
\end{eqnarray}
Next, we express $P_J(\cos(2\theta))$ as
\begin{equation}
P_J(\cos(2\theta))=\sum_{L=0}^{\infty}d_{L,J}P_L(\cos\theta)\label{dLJ},
\end{equation}
where the coefficients $d_{L,J}$ are generated by an efficient
recurrent procedure described in Appendix A of Ref.
\cite{LeibscherOther}. Using the well-known relation between
Legendre polynomials and spherical harmonics, one finds
$\alpha_l^{m_0}$ to be
\begin{eqnarray}
\alpha_l^{m_0}&=&\exp[-i\frac{|P_s|}{2}]\sum_{J=0}^{\infty}\sum_{L=0}^{\infty}d_{L,J}i^{3J}(2J+1)\nonumber\\
&\times&j_J(\frac{|P_s|}{2})\sqrt{\frac{2l_0+1}{2l+1}}\nonumber\\
&\times&C(L,l_0,l|0,0,0)C(L,l_0,l|0,m_0,m_0)\label{alphalm0},
\end{eqnarray}
where $C(L,l_0,l|0,0,0), C(L,l_0,l|0,m_0,m_0)$ are Clebsch-Gordan
coefficients. Free evolution  after the first pulse gives
\begin{equation}
\psi_{l_0}^{m_0}(t_1^-)=\sum_{l_0=0}^{\infty}\alpha_{l}^{m_0}exp[-\frac{i}{2}l(l+1)t_1]|l,m_0\rangle.
\label{FirstProp}
\end{equation}
Next, we apply the second (asymmetric) pulse at time $t_1$,
\begin{equation}
\psi_{l_0}^{m_0}(t_1^+)=\exp[iP_a\cos(\theta)]\psi_{l_0}^{m_0}(t_1^-)\label{dipole111},
\end{equation}
and again re-expand the new wave function,
\begin{equation}
\psi_{l_0}^{m_0}(t_1^+)=\sum_{\bar{l}=0}^{\infty}\tilde{\alpha}_{\bar{l}}^{m_0}|\bar{l},m_0\rangle
\label{FirstExpansion111},
\end{equation}
in order to find the spherical harmonic coefficients:
\begin{eqnarray}
\tilde{\alpha}_{\bar{l}}^{m_0}&=&\sum_{J=0}^{\infty}\sum_{l=0}^{\infty}i^Jj_J(P_a)(2J+1)\alpha_l^{m_0}\exp[-il(l+1)\frac{t_1}{2}]\nonumber\\
&\times&\sqrt{\frac{2l+1}{2\bar{l}+1}}C(J,l,\bar{l}|0,0,0)C(J,l,\bar{l}|0,m_0,m_0).\label{Secondalphaterm}
\end{eqnarray}
Finally, this new wave function propagates freely for a time
interval $t_2$, giving
\begin{equation}
\psi_{l_0}^{m_0}(t=t_1+t_2)=\sum_{\bar{l}=0}^{\infty}\tilde{\alpha}_{\bar{l}}^{m_0}\exp[-\frac{i}{2}\bar{l}(\bar{l}+1)t_2]|\bar{l},m_0\rangle.
\label{SecondProp}
\end{equation}
This last expression is substituted into Eq. (\ref{Alignmentl0m0})
in order to calculate the orientation factor given by Eq.
(\ref{Alignment}).

\bibliographystyle{phaip}

\end{document}